# A Gas-Phase Kinetic Study of the N($^2$D) + CH$_3$CCH and N($^2$D) + CH$_3$CN Reactions

Kevin M. Hickson,[a] Jean-Christophe Loison,[a] Benjamin Benne[b,c] and Michel Dobrijevic[d]

[a] *Univ. Bordeaux, CNRS, Bordeaux INP, ISM, UMR 5255, F-33400 Talence, France*
[b] *The University of Edinburgh, School of GeoSciences, Edinburgh, UK*
[c] *Centre for Exoplanet Science, University of Edinburgh, Edinburgh, UK*
[d] *Univ. Bordeaux, CNRS, LAB, UMR 5804, F-33600 Pessac, France*

**Abstract**

The chemistry of planetary atmospheres containing molecular nitrogen as a major atmospheric component is strongly influenced by the reactions of atomic nitrogen. Although nitrogen atoms in their ground electronic state N($^4$S) are mostly unreactive towards stable molecules, electronically excited nitrogen atoms N($^2$D) are much more reactive and could play an important role in the formation of nitriles and other nitrogen bearing organic molecules in planetary atmospheres such as Titan. Despite this, few kinetic studies of N($^2$D) reactions have been performed over the appropriate low temperature range. Here, we report the results of an experimental study of the reactions N($^2$D) + methylacetylene, CH$_3$CCH, and N($^2$D) + acetonitrile, CH$_3$CN, using a supersonic flow reactor at selected temperatures between 50 K and 296 K. N($^2$D) atoms, which were generated indirectly as a product of the C($^3$P) + NO reaction, were subsequently detected by laser induced fluorescence in the vacuum ultraviolet wavelength region. The measured rate constants are significantly larger than the estimated values in current photochemical models and do not display large variations as a function of temperature. The new rate constants are included in a 1D coupled ion-neutral model of Titan's atmosphere to test their influence on the simulated species abundances. In addition, the overall description of both reactions is improved by considering the results of recent experimental and theoretical work examining the product channels of these processes. These simulations indicate that while the N($^2$D) + CH$_3$CCH reaction has only a limited overall influence on Titan's atmospheric chemistry, the N($^2$D) + CH$_3$CN reaction could lead to the formation of significant relative abundances of cyanomethamine, HNCHCN, in the upper atmosphere.

# 1 Introduction

Nitrogen is among the most abundant of all elements, while compounds bearing elemental nitrogen are omnipresent throughout the Universe. In its atomic form, ground electronic state nitrogen, N($^4$S) is characterized by a low inherent reactivity towards molecules with a full valence shell so that its reactions with radical species are the most important mechanisms for incorporating N($^4$S) atoms in molecules at room temperature and below. [1,2,3,4,5,6,7,8]

In contrast to N($^4$S), atomic nitrogen in its first excited electronic state, N($^2$D), has been shown in kinetic studies to react much more rapidly with closed shell molecules. [9,10] In environments where molecular nitrogen, N$_2$ is present at high abundance levels, such as in the atmospheres of the Earth, Pluto and Saturn's moon Titan, N($^2$D) atoms are formed in the upper atmosphere by the photodissociation of N$_2$ at vacuum ultraviolet wavelengths below 102 nm. [11] As N($^2$D) atoms are metastable with long radiative lifetimes (13.6 and 36.7 hours for the $^2$D$_{3/2}$ and $^2$D$_{5/2}$ fine structure states respectively), [12] and are only slowly quenched by N$_2$, [13,14] these atoms could make an important contribution to the chemistry of planetary atmospheres. In Titan's atmosphere in particular, which is composed of molecular nitrogen (95 %) and methane, CH$_4$, present at the 2-5 % level and other larger organic species in trace amounts, the reactions between N($^2$D) atoms and organic molecules could be important pathways for the formation of more complex nitrogen bearing species such as amines, nitriles and imines. Indeed, several reactions involving N($^2$D) atoms were identified as key processes in earlier photochemical modeling studies of Titan's atmosphere. [15,16,17] Previous dynamics studies of the reactions of N($^2$D) atoms with a range of organic molecules have already demonstrated the relevance of these processes in the formation of a wide range of nitrogen bearing organic molecules, [18,19,20,21,22,23] while recent kinetics studies [21,23,24,25,26,27] have measured the rate constants for certain N($^2$D) reactions over the appropriate low temperature range for the first time. Given their potential relevance to the overall chemistry of Titan's atmosphere, it is important to characterize the reactions of N($^2$D) with the most abundant atmospheric species in terms of both their rate constants as a function of temperature as well as their product formation channels. In this respect, methylacetylene (propyne) CH$_3$CCH, and acetonitrile (methyl cyanide) CH$_3$CN have both been detected and are relatively abundant molecules in Titan's atmosphere. Observations by Lombardo et al. [28] derived mixing ratios higher than 10$^{-8}$ for CH$_3$CCH, while for CH$_3$CN, Thelen et al. [29] derived mixing ratios increasing from a few 10$^{-9}$ at 200 km to 10$^{-7}$ at 500 km. Earlier measurements by Cui et al. [30] using the Ion Neutral Mass Spectrometer (INMS) instrument of Cassini at 1000 km determined values as high as 10$^{-6}$ for the CH$_3$CN mixing ratio. Indeed, as photochemical models predict that the mole fractions of both CH$_3$CCH and CH$_3$CN reach their peak around 1000 km, [17] at similar altitudes to the simulated production peak for N($^2$D) atoms, the reactions of these molecules with N($^2$D) could make a significant contribution to the production of N-bearing molecules in Titan's atmosphere.

To this end, we describe an experimental kinetics study of the N($^2$D) + CH$_3$CCH and N($^2$D) + CH$_3$CN reactions using a supersonic flow reactor over the 50-296 K temperature range (the N($^2$D) + CH$_3$CN reaction was only studied down to 75 K). In a similar manner to our previous studies of N($^2$D) reactions, N($^2$D) atoms were produced by chemical reaction and were detected by laser induced fluorescence at vacuum ultraviolet wavelengths (VUV LIF). The rate constants for these reactions were introduced into an updated photochemical model of Titan's atmosphere to test their effects on species abundances. This paper is organized as follows. The experimental methods are presented in section 2, while the results are presented and discussed in section 3. Section 4 describes the photochemical model and discusses the main effects brought about by the new rate constants. The major conclusions of this work are summarized in section 5.

## 2 Experimental methods

An existing supersonic flow reactor, a technique also known by the acronym CRESU (Cinétique de Réaction en Écoulement Supersonique Uniforme) was used to conduct the present measurements. Its major design features were reported in the earliest papers performed with this apparatus. [31][32] The original system has undergone significant modifications over the years to allow us to investigate gas-phase reactions involving atomic radicals; species that typically only possess strong electronic transitions in the vacuum ultraviolet (VUV) wavelength region. Through the development and implementation of a method to produce narrowband tunable radiation in the 115-128 nm region, it has been possible to study the reactions of both ground state atoms (C($^3$P), [33][34] H($^2$S)/D($^2$S), [35][36] N($^4$S) [8]) and excited state atoms (O($^1$D), [37][38] N($^2$D), [39][24]) from room temperature down to 50 K. With regard to the present investigation, the electronic quenching of N($^2$D) by Ar[40] and N$_2$[14] is inefficient, allowing us to employ both of these carrier gases in the supersonic flow. Three Laval nozzles were used during this work, although as one of the nozzles was used with both carrier gases, it was possible to generate flows with four different characteristic temperatures between 50 and 177 K. The flow properties are summarized in Table 1 of Nunez-Reyes et al. [25] In order to access room temperature, the apparatus was also used without the Laval nozzle and at reduced flow velocities. Under these conditions, the gas in the interaction region was replenished between laser shots. As N($^2$D) atoms are difficult to produce photolytically with standard laser sources, an alternative method was adopted to generate this reagent. Here, N($^2$D) atoms were formed as the product of the gas-phase reaction between ground state atomic carbon C($^3$P) and nitric oxide NO

$$C(^3P) + NO \rightarrow N(^2D, ^4S) + CO \quad (1a)$$
$$\rightarrow O(^3P) + CN \quad (1b)$$

following the method described by Nunez-Reyes & Hickson.[39] According to earlier work, the rate constant for channel (1a) is larger than the one for channel (1b) with an estimated branching ratio [N($^2$D) + N($^4$S)]/[O($^3$P)] of 1.5 ± 0.3 at 298 K.[41] C($^3$P) atoms were produced by pulsed laser photolysis of the precursor molecule tetrabromomethane (CBr$_4$) at 266 nm. The pulse energy was typically around 38 mJ with a 5 mm diameter beam. CBr$_4$, a crystalline solid at room temperature, was carried into the reactor by diverting some of the main carrier gas flow into a flask containing CBr$_4$. The gas flow into the flask was constant for any single series of experiments in addition to the flask pressure and temperature so that the gas-phase CBr$_4$ concentration did not vary. The maximum CBr$_4$ concentration was estimated to be approximately 4 × 10$^{13}$ cm$^{-3}$ based on its saturated vapour pressure.

A known coproduct of CBr$_4$ photolysis at 266 nm is atomic carbon in its first singlet excited state, C($^1$D). According to the earlier work of Shannon et al.[42] performed under similar conditions, the ratio C($^1$D)/C($^3$P) = 0.1-0.15 for this multiphoton dissociation process. Possible interferences from the reactions of C($^1$D) atoms are discussed in section 3.

N($^2$D) atoms were detected by pulsed laser induced fluorescence (LIF) in the VUV wavelength range in the present work (VUV LIF). Here, narrowband tuneable radiation at 116.745 nm was used to excite the 2s$^2$2p$^3$ $^2$D° - 2s$^2$2p$^2$($^3$P)3d $^2$F transition, with the fluorescence emission detected on-resonance. Tunable light around 116.7 nm was generated in a two-step process. First, the 700.5 nm output of a pulsed tuneable dye laser was steered into a BBO crystal for frequency doubling purposes, generating a beam of UV radiation at 350.25 nm with a pulse energy of approximately 9 mJ. Second, after the residual fundamental radiation was discarded through the use of two dichroic mirrors centred on 355 nm, the resulting UV beam was directed and focused into a cell containing 40 Torr of Xe and 560 Torr of Ar, generating tunable VUV radiation by frequency tripling. A magnesium fluoride (MgF$_2$) lens acted as the tripling cell exit window, allowing the divergent VUV beam to be collimated and steered into the reactor. The tripling cell was attached to the reactor via a long (75 cm) sidearm to prevent atmospheric absorption of the VUV beam. A series of circular diaphragms were placed along the length of the sidearm to capture the divergent residual UV radiation before it could reach the reactor. Upon entering the reactor, the VUV beam crossed the supersonic flow at right angles, exciting unreacted N($^2$D) atoms. The resonant VUV LIF emission from these atoms was collected by a lithium fluoride (LiF) lens placed at 90° to the plane containing both the supersonic flow and the VUV excitation laser to minimize the detection of scattered light. This lens, which focused the emission onto the photocathode of a solar blind photomultiplier tube (PMT), was protected from the gases in the reaction chamber by a LiF window. The isolated region between the PMT and the LiF window was continuously evacuated to maximize transmission of the VUV light. The PMT output signal was amplified and fed into a gated integration system. VUV intensities were recorded as a function of reaction time which corresponded to the delay time between the photolysis and probe laser pulses. This delay was precisely controlled by a digital pulse generator. Data was acquired for at least 70 different delay times with each intensity value the average of 30 probe laser shots at each delay time. The zero signal baseline level was established by recording several delay times at negative values; that is, with the probe laser firing before the photolysis laser. The gases used in the experiments (Linde Ar 99.999%, Xe 99.999%, Air Liquide CH$_3$CCH ≥ 96%, Air Liquide N$_2$ 99.999%, NO 99.9%) were flowed directly from the cylinders into calibrated mass-flow controllers, which were used to precisely regulate gas flows into the reactor. Liquid CH$_3$CN was purchased as a commercial reagent (≥ 99.9 %) and was not purified prior to use.

## 3 Results and discussion

The concentrations of NO, [NO], and either CH$_3$CN, [CH$_3$CN] or CH$_3$CCH, [CH$_3$CCH] were maintained in large excess with respect to the concentrations of C($^3$P) (as the precursor for N($^2$D) formation) and N($^2$D) for all the experiments reported here. Consequently, the concentrations of these coreagent species were effectively constant for any individual experiment thereby allowing a first-order analysis to be applied (the so-called pseudo-first-order approximation). As a result, the N($^2$D) fluorescence signal obeyed a biexponential temporal profile of the form

$$I_{N(^2D)} = A(\exp(-k'_a t) - \exp(-k'_b t)) \qquad (2)$$

where $k'_a$ is the pseudo-first-order rate constant for N($^2$D) loss, $k'_b$ is the pseudo-first-order rate constant for N($^2$D) formation, A is the maximum signal amplitude in the absence of competing losses (ie when $k'_a = 0$) and $t$ is time. The contributions to the constants denoted by $k'_a$ and $k'_b$ are described in Nunez-Reyes et al.,[23] with either $CH_3CN$ or $CH_3CCH$ in the place of $C_2H_2$. In common with this earlier work, a single exponential function of the form

$$I_{N(^2D)} = A\exp(-k'_a t) \qquad (3)$$

was actually used to fit the fluorescence signal profiles due to the impossibility to exploit the first few microseconds following the photolysis laser pulse due to amplifier saturation issues. The starting point of the fit was adjusted to begin only after $N(^2D)$ production by reaction (3a) was essentially complete. In practice, this was verified by plotting the logarithm of the VUV LIF signal as a function of time and by exploiting only the 'linear' part of the decay profile. Two representative $N(^2D)$ fluorescence profiles recorded at 50 K during our experiments on the $N(^2D) + CH_3CCH$ reaction are shown in Figure 1.

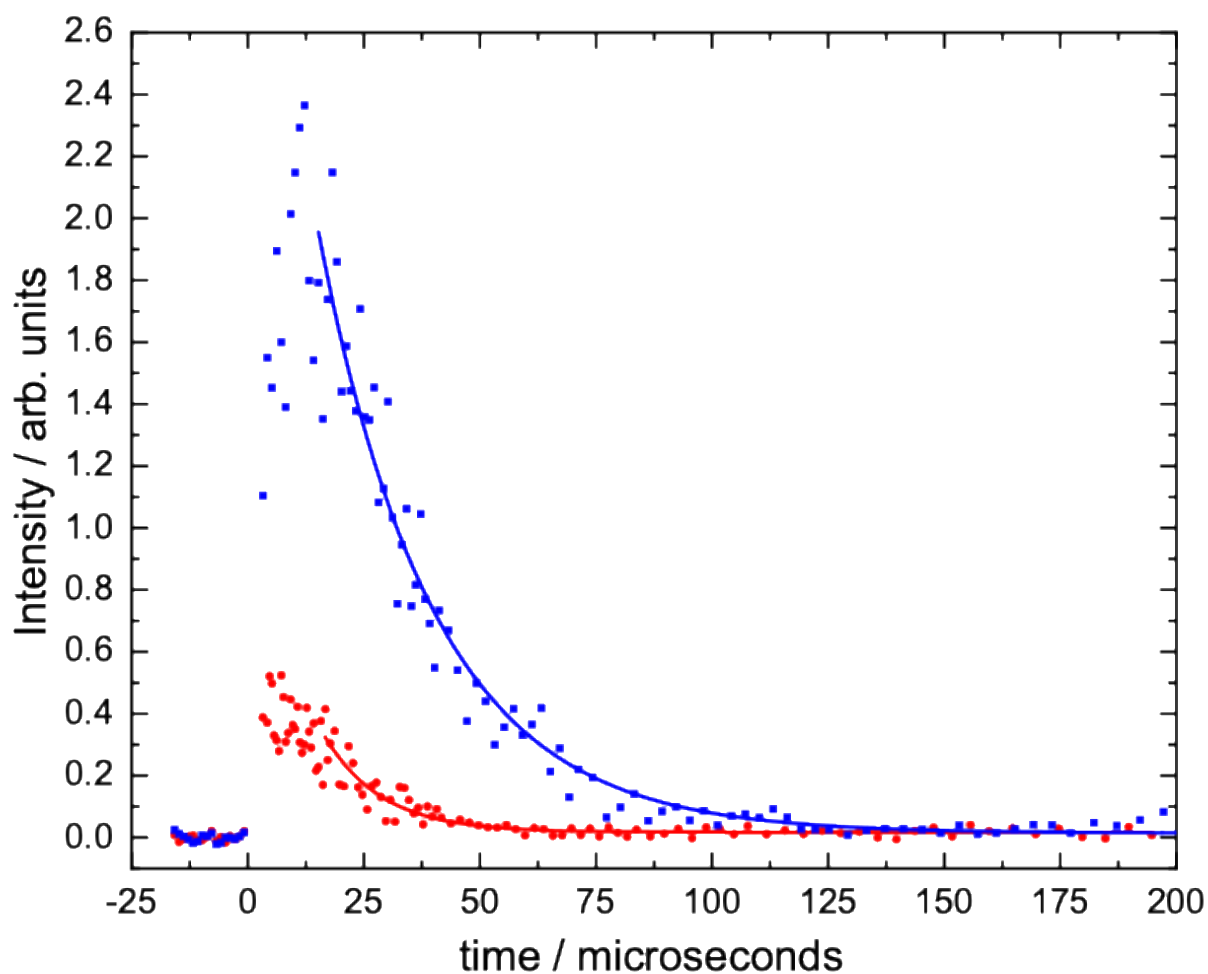

**Figure 1** $N(^2D)$ fluorescence intensity as a function of reaction time recorded at 50 K. (Solid blue squares) without $CH_3CCH$; (solid red circles) $[CH_3CCH] = 2.5 \times 10^{14}$ $cm^{-3}$. $[NO] = 4.3 \times 10^{14}$ $cm^{-3}$ for both decays. Single exponential fits to the data according to expression (3) are represented by solid red and blue lines.

**Effects of the competing C($^3$P) + CH$_3$CN and C($^3$P) + CH$_3$CCH reactions**

As atomic carbon (in addition to N($^2$D)) is expected to react rapidly with both CH$_3$CN[43] and CH$_3$CCH[44] in the present experiments, we need to consider any possible interferences that these reactions might induce.

The most obvious effect of these processes is the decrease in N($^2$D) production, clearly seen in Figure 1. Here, atomic carbon which would be available to form N($^2$D) atoms through reaction (3a) in the absence of CH$_3$CN or CH$_3$CCH is instead removed from the supersonic flow by the competing reactions with these reagents. Nevertheless, although the signal-to-noise ratio is lower for those experiments performed at high coreagent concentrations, it is still high enough to yield accurate fits.

In addition to the effects of these reactions on the relative N($^2$D) signal levels, we also need to evaluate whether their products could affect our kinetic study of the N($^2$D) + CH$_3$CN/CH$_3$CCH reactions by themselves reacting with excess reagents NO and CH$_3$CN or CH$_3$CCH. In this respect, it is important to check that additional N($^2$D) atoms are not produced by secondary chemistry. Previous studies of the C($^3$P) + CH$_3$CN reaction indicate that the major reaction products should be H + CH$_2$CNC and/or H + CH$_2$CCN over the triplet energy surface with the possible formation of C$_2$H$_2$ + HCN/HNC and/or H$_2$ + HC$_3$N through intersystem crossing to the singlet potential energy surface (PES).[43] None of the products formed over the singlet PES will react further with CH$_3$CN, while it is not immediately obvious how the reactions between CH$_2$CNC/CH$_2$CCN and CH$_3$CN or NO could lead to additional N($^2$D) formation. In the case of the C($^3$P) + CH$_3$CCH reaction, reaction occurs by addition to the C-C triple bond followed by H-atom elimination to form H$_2$CCCCH + H as products.[45] As before, it also seems very unlikely that the reaction of H$_2$CCCCH with NO could lead to additional N($^2$D) formation. Indeed, a previous experimental and theoretical study of the related C$_2$H + NO reaction[46] determined that the major product channels were HCN + CO and HCO + CN rather than the formation of atomic products.

With regard to the presence of C($^1$D) (produced by CBr$_4$ photolysis) in the supersonic flow, the possible effects of several secondary processes including the reaction of C($^1$D) with NO have already been considered in earlier work.[27,39]

Although no information exists in the literature regarding the C($^1$D) + CH$_3$CN reaction, it is expected that this process is rapid, considering the large measured rate constants for the C($^3$P) + CH$_3$CN reaction.[43] A previous theoretical study of the related reaction between O($^1$D) atoms[47] and CH$_3$CN predicts that the C-N triple bond remains intact with the formation of CH$_3$ + NCO, CH$_2$CN + OH, H + HOCHCN and H$_2$O + HCCN as the major product channels, so this process is highly unlikely to lead to the formation of additional N($^2$D) atoms. Similarly, although there are no previous kinetics studies of the C($^1$D) +CH$_3$CCH reaction, this process should also be rapid by comparison with the C($^3$P) + CH$_3$CCH reaction,[48] with predicted major products $i$-C$_4$H$_3$ + H, C$_2$H$_2$ + CCH$_2$ and C$_4$H$_2$ + H$_2$.[49]

During these experiments, no fewer than 14 separate decays similar to those shown in Figure 1 were recorded at any given temperature (with as many as 47), over a range of coreagent (CH$_3$CN or CH$_3$CCH) concentrations. During our investigation of the N($^2$D) + CH$_3$CN reaction in particular, the range of useable CH$_3$CN concentrations was very limited at low temperature due to the formation of CH$_3$CN clusters in the flow. Similar problems were also encountered in earlier studies of CH$_3$CN reactions.[43,50] Consequently, it was not possible to make a reliable measurement of the rate constant for this reaction at 50 K. In the present experiments, where N($^2$D) atoms are also lost through the competing N($^2$D) + NO reaction, the change in the value of $k_{1st}$ brought about by the addition of CH$_3$CN to the flow can be quite small compared to the y-intercept value of the second-order plot (representing N($^2$D) loss by the N($^2$D) + NO reaction alone). At 75 K, the contribution to $k_{1st}$ due to the N($^2$D) + NO reaction alone is approximately 26500 s$^{-1}$ following the previous work of Nunez-Reyes et al.[39] With [CH$_3$CN] limited to values less than $1.9 \times 10^{13}$ cm$^{-3}$ at this temperature, the contribution to the overall loss rate of N($^2$D) is approximately 10 % of the total, so the precision on the measured rate constant value is quite low. Figure 2 shows the variation of the pseudo-first-order rate constant as a function of the coreagent concentration (panel A - CH$_3$CCH , panel B - CH$_3$CN), where the pseudo-first-order rate constants were derived from fits to temporal profiles such as the ones shown in Figure 1 using expression (3).

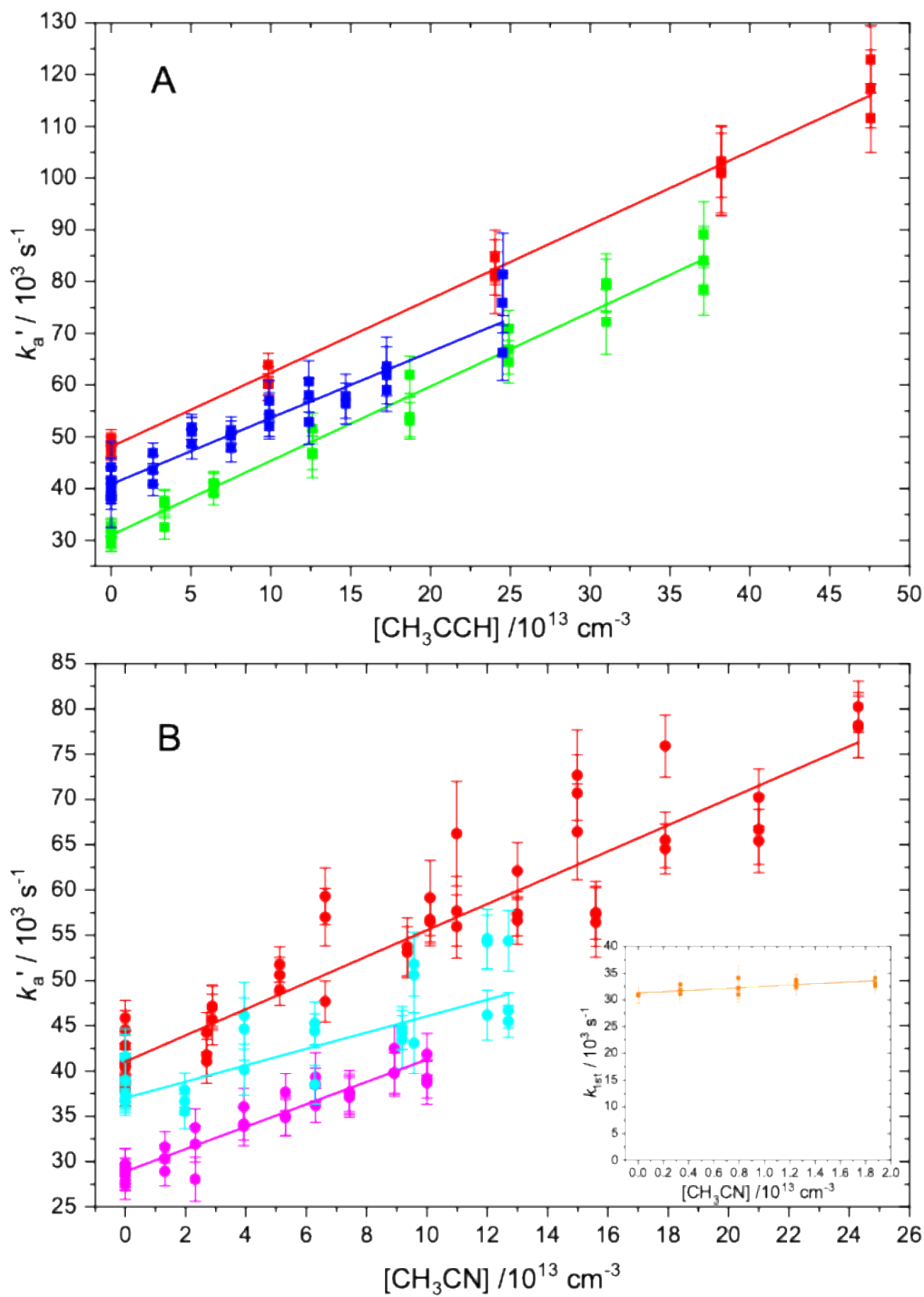

**Figure 2** Derived pseudo-first-order rate constants $k'_a$ as a function of the coreagent concentration. Panel A - The N($^2$D) + CH$_3$CCH reaction. (Red solid squares) 296 K, N$_2$ carrier gas; (green solid squares) 75 K; (blue solid squares) 50 K. Panel B - The N($^2$D) + CH$_3$CN

reaction. (Red solid circles) 296 K, Ar carrier gas; (magenta solid circles) 177 K; (cyan solid circles) 127 K. Inset – (orange solid squares) 75 K. Solid lines depict weighted linear least-squares fits to the data. The error bars on individual data points are shown at the level of a single standard deviation and are derived from exponential fits using expression (3) of the time dependent intensity profiles displayed in Figure 1.

The second-order rate constants, $k_{2nd}$, were then obtained from the slopes of weighted linear-least squares fits to the data as shown by the solid lines in Figure 2. The second-order rate constants obtained are listed in Table 1 alongside other relevant information. Figure 3 shows the second-order rate constants for both reactions plotted as a function of temperature.

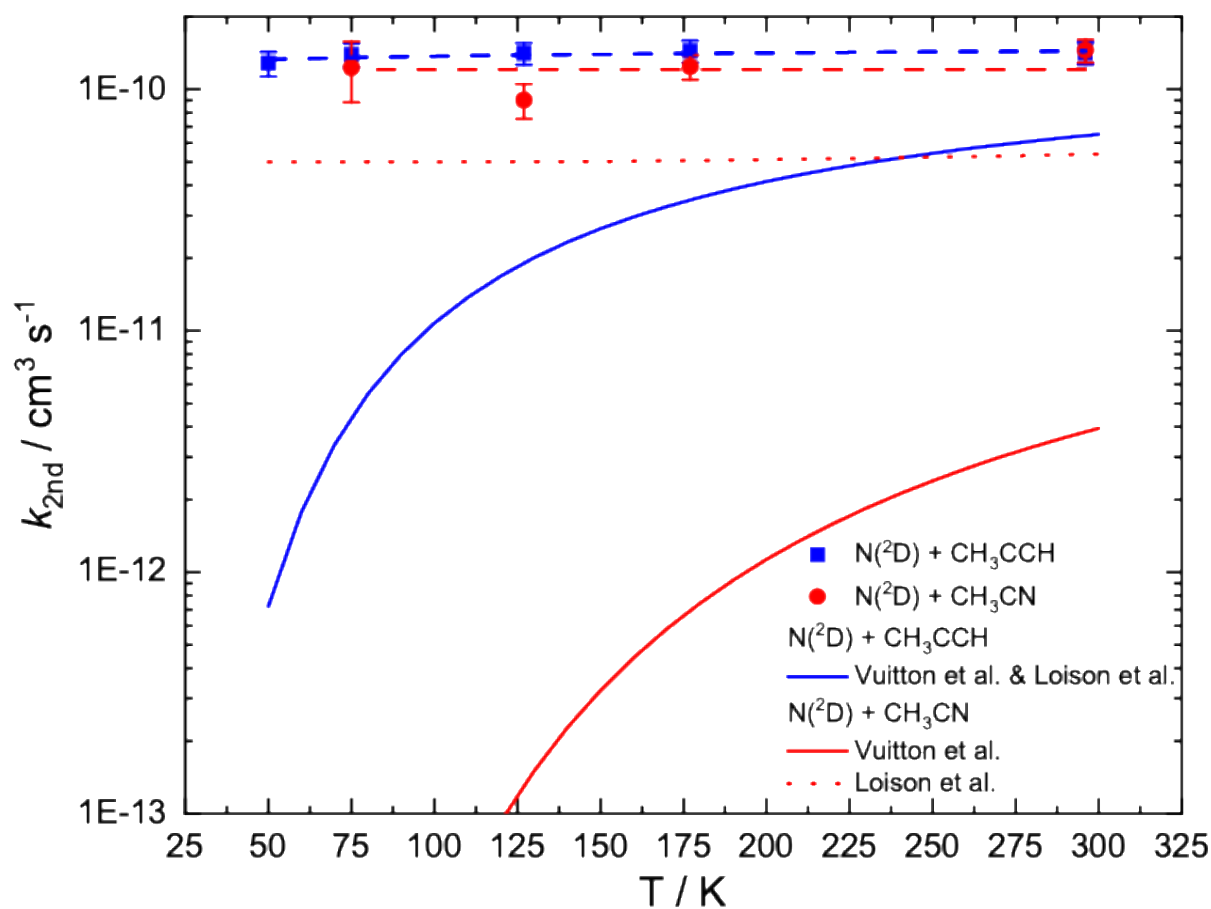

**Figure 3** Second-order rate constants for the N($^2$D) + CH$_3$CCH and N($^2$D) + CH$_3$CN reactions as a function of temperature. (Blue solid squares) the N($^2$D) + CH$_3$CCH reaction; (red solid circles) the N($^2$D) + CH$_3$CN reaction. Error bars on the present values represent the statistical (1σ) and systematic uncertainties (estimated to be 10 %). (Dashed blue line) temperature dependent fit to the N($^2$D) + CH$_3$CCH reaction rate constants. (Dashed red line) temperature independent value for the N($^2$D) + CH$_3$CN reaction rate constants. (Solid blue line) Rate

constant values for the N($^2$D) + CH$_3$CCH reaction used in the Vuitton et al.[51] and Loison et al.[17] photochemical models; (Solid red line) Rate constant values for the N($^2$D) + CH$_3$CN reaction used in the Vuitton et al. photochemical model.[51] (Dotted red line) Rate constant values for the N($^2$D) + CH$_3$CN reaction used in the Loison et al. photochemical model.[17]

**Table 1** Measured $k_{2nd}$ values for the N($^2$D) + CH$_3$CCH and N($^2$D) + CH$_3$CN reactions

| T / K | N$^b$ | Flow density / 10$^{16}$ cm$^{-3}$ | [CH$_3$CCH] / 10$^{14}$ cm$^{-3}$ | [NO] / 10$^{14}$ cm$^{-3}$ | $k_{N(^2D)+CH_3CCH}$ / 10$^{-11}$ cm$^3$ s$^{-1}$ |
|---|---|---|---|---|---|
| 296 (Ar) | 27 | 16.3 | 0 – 4.8 | 6.3 | (14.1 ± 1.5)$^c$ |
| 296 (N$_2$) | 15 | 16.3 | 0 – 4.8 | 6.2 | (14.3 ± 1.5) |
| 177 ± 2$^a$ | 27 | 9.4 ± 0.2$^a$ | 0 – 3.7 | 4.1 | (14.4 ± 1.5) |
| 127 ± 2 | 27 | 12.6 ± 0.3 | 0 – 4.2 | 4.6 | (14.1 ± 1.5) |
| 75 ± 2 | 27 | 14.7 ± 0.6 | 0 – 2.1 | 3.0 | (13.9 ± 1.6) |
| 50 ± 1 | 32 | 25.9 ± 0.9 | 0 – 1.7 | 4.3 | (12.8 ± 1.5) |
| T / K | N$^b$ | | [CH$_3$CN] / 10$^{13}$ cm$^{-3}$ | [NO] / 10$^{14}$ cm$^{-3}$ | $k_{N(^2D)+CH_3CN}$ / 10$^{-11}$ cm$^3$ s$^{-1}$ |
| 296 | 47 | 16.3 | 0 – 24.3 | 6.3 | (14.5 ± 1.6) |
| 177 ± 2 | 30 | 9.4 ± 0.2 | 0 – 10.0 | 4.4 | (12.4 ± 1.5) |
| 127 ± 2 | 27 | 12.6 ± 0.3 | 0 – 12.7 | 4.7 | (9.0 ± 1.5) |
| 75 ± 2 | 14 | 14.7 ± 0.6 | 0 – 1.9 | 3.0 | (12.3 ± 3.5) |

$^a$Uncertainties on the calculated temperatures and flow densities represent the statistical (1σ) errors obtained from Pitot tube measurements of the impact pressure. $^b$Number of individual measurements. $^c$Uncertainties on the measured rate constants represent the combined statistical (1σ) and estimated systematic (10%) errors.

Although neither of the N($^2$D) + CH$_3$CCH and N($^2$D) + CH$_3$CN reactions are currently included in astrochemical databases such as the Kinetic Database for Astrochemistry (KIDA),[52] they have been included in photochemical models of Titan's atmosphere, based on estimated parameters. Loison et al.[17] considered the following channels for this process;

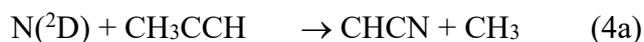

N($^2$D) + CH$_3$CCH  → CHCN + CH$_3$     (4a)

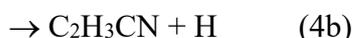

→ C$_2$H$_3$CN + H     (4b)

with a total rate constant varying between 6.5 × 10$^{-11}$ cm$^3$ s$^{-1}$ at 300 K to 3.4 × 10$^{-12}$ cm$^3$ s$^{-1}$ at 70 K, split equally between the two channels. The photochemical model of Vuitton et al.[51] assumed a single reactive channel leading to C$_3$H$_3$N + H as products, with the same rate constant values as the ones adopted by Loison et al.[17] A recent experimental and theoretical study of the N($^2$D) + CH$_3$CCH reaction by Mancini et al.[53] has shed more light on the preferred product channels of this process. Here, a crossed molecular beam study coupled with mass selected product detection by electron ionization was used to probe the products of collisions at mass-to-charge ratios (m/z) of 53, 52 and 51 between a beam of N($^2$D) atoms and a beam of

CH₃CCH molecules at a collision energy of 31 kJ/mol. In addition, Mancini et al. [53] also performed a detailed study of the PES of the N($^2$D) + CH₃CCH reaction, coupled with Rice Ramsperger Kassel Marcus (RRKM) calculations to obtain information on the various product formation channels arising from unimolecular dissociation of the intermediate species. Theoretically, they determined that reaction should proceed by two main mechanisms. The first one occurs via a barrierless addition of the N($^2$D) atom to the C-C triple bond of CH₃CCH, while the second one occurs through insertion of the N($^2$D) atom into the C-H bond of the methyl group. According to their RRKM calculations, the product channels CH₂NH + C₂H, c-C(N)CH + CH₃ and CH₂CHCN + H were found to be the major product channels (with branching ratios of 0.41, 0.32 and 0.12 respectively) while the experiments themselves were only able to derive information about the H-atom production channels. Considering the barrierless nature of the N($^2$D) + CH₃CCH PES and a recent study of the kinetics of the closely related N($^2$D) + CH₂CCH₂ reaction down to low temperatures which determined large rate constants for this process (1.7 × 10⁻¹⁰ cm³ s⁻¹ independent of temperature over the 50-296 K range), Mancini et al. hypothesized that the rate constant for the N($^2$D) + CH₃CCH should be considerably larger than the values used in current photochemical models.

The photochemical model of Loison et al. [17] considers three channels for the N($^2$D) + CH₃CN reaction, namely;

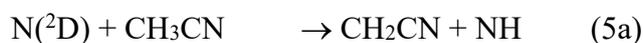
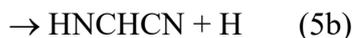
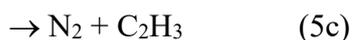

$$N(^2D) + CH_3CN \rightarrow CH_2CN + NH \quad (5a)$$
$$\rightarrow HNCHCN + H \quad (5b)$$
$$\rightarrow N_2 + C_2H_3 \quad (5c)$$

with a total rate constant close to 5 × 10⁻¹¹ cm³ s⁻¹ over the relevant temperature range (reactions (5a) and (5b) account for only a few percent of the overall products). Vuitton et al. [51] considered only channels (5a) and (5b) in their photochemical model (with HNCHCN considered as the generic C₂H₂N₂ species) with an overall rate constant ranging from 4 × 10⁻¹² cm³ s⁻¹ at 300 K to 1 × 10⁻¹⁵ cm³ s⁻¹ at 70 K, a significantly smaller value than the one estimated by Loison et al. [17] To the best of our knowledge however, there are no earlier experimental studies of the reaction between N($^2$D) and CH₃CN. On the theoretical side, Mancini et al. [54] recently performed a detailed quantum chemical characterization of the reactive PES of the N($^2$D) + CH₃CN system. In addition to the PES calculations, they also performed RRKM estimates of the exit channels to determine the preferred reaction products. They showed that the major reaction pathway should be N($^2$D) insertion into one of the C-H bonds of the CH₃CN methyl group leading the formation of HNCHCN in cis- and trans- conformers and a H-atom

(equivalent of reaction (5b)). Other predicted minor products include the H + $CH_2NCN$ channel at the 5 % level. Channel (5c) formed by the attack of N($^2$D) on the C-N triple bond of $CH_3CN$ or on the nitrogen atom lone pair was found to be negligible despite it being the most exothermic exit channel (-459 kJ/mol at the CCSD(T)/aug-cc-pVTZ level). In terms of the reaction rate, by analogy with other similar reactions of N($^2$D) with species containing methyl groups such as $CH_4$ and $C_2H_6$ they recommended a value in the $10^{-11}$ $cm^3$ $s^{-1}$ range for the total rate constant of reaction (5).

Considering our experimental results for the N($^2$D) + $CH_3CCH$ reaction, if we fit these data using a modified Arrhenius expression of the form k($T$) = α ($T$/300)$^β$ exp(-γ/$T$) to represent the magnitude of the rate constant for inclusion in photochemical models, we obtain the parameters α = (14.4 ± 1.3) × $10^{-11}$ $cm^3$ $s^{-1}$ and β = 0.05 ± 0.02 (with γ = 0). This equates to a rate constant value that is nearly constant over the 50-296 K range, with predicted values of $k_{N(^2D)+CH_3CCH}$(296 K) = 14.4 × $10^{-11}$ $cm^3$ $s^{-1}$ and $k_{N(^2D)+CH_3CCH}$(50 K) = 13.2 × $10^{-11}$ $cm^3$ $s^{-1}$. For the case of the N($^2$D) + $CH_3CN$ reaction, given the relatively scattered nature of the data, a temperature independent value for the rate constant of (12.0 ± 2.3) × $10^{-11}$ $cm^3$ $s^{-1}$ is preferred.

Assuming an average temperature of 170 K for the atmosphere of Titan leads to a predicted rate constant for the N($^2$D) + $CH_3CCH$ reaction $k_{N(^2D)+CH_3CCH}$(170 K) = 14.0 × $10^{-11}$ $cm^3$ $s^{-1}$ that is more than four times larger than the value estimated by Vuitton et al. [51] and Loison et al. [17] of 3.3 × $10^{-11}$ $cm^3$ $s^{-1}$. For the N($^2$D) + $CH_3CN$ reaction where the reactivity between these two species was estimated by comparison with the N($^2$D) + $CH_4$ reaction which is characterized by an activation barrier, the measured rate constant for this process is more than two hundred times larger than the value estimated by Vuitton et al. [51] The values estimated by Loison et al. [17] for the N($^2$D) + $CH_3CN$ reaction of $k_{N(^2D)+CH_3CN}$(170 K) = 5.1 × $10^{-11}$ $cm^3$ $s^{-1}$ are closer to the measured ones, with the experimental rate constants being more than two times larger at 170 K. The parameters for the rate constant expressions and the various product channels of these reactions used in the photochemical model can be found in Table S1 of the supplementary information file.

**5 Photochemical model**

In order to test the effect of these measurements on the chemistry of Titan's atmosphere, we included the new rate constants for the N($^2$D) + $CH_3CCH$ and N($^2$D) + $CH_3CN$ reactions in a photochemical model. The 1D-model of Dobrijevic et al. [55] was used as the basis for this work, which treats the coupled chemistry of neutrals and cations from the lower atmosphere to the ionosphere with updated chemistry of aromatic compounds,[56] nitrogen bearing molecules[15, 17] as well as various N($^2$D) reactions[24, 25, 27] including our recent study of the N($^2$D) + allene reaction.[21] This model also includes the transport of magnetospheric electrons as described in Benne et al.[57] The new product channels of the N($^2$D) + CH$_3$CCH and N($^2$D) + CH$_3$CN reactions have been deduced from the theoretical work of Mancini et al.[53, 54] For the N($^2$D) + CH$_3$CCH reaction, we have not introduced any new product species in the network because the only new species, according to Mancini et al.,[53] is c-C(N)CH that we consider to be HCCN in this work. These two radicals will in fact have similar reactivity, in particular they will react with H and CH$_3$ to give similar products. Furthermore, since HCCN is produced by the reactions CH + HCN and N($^2$D) + C$_2$H$_2$ with fluxes well above those produced by the N($^2$D) + CH$_3$CCH reaction, the contribution of the latter reaction will not distort the model. For the N($^2$D) + CH$_3$CN reaction we have introduced a new product channel as the main products are calculated to be HNCHCN + H.[54] This species was already present in the network, but we have added its most important loss processes (including photodissociation calculated at the EOM-CCSD/AVTZ level, see the electronic supplementary information, considering that its photodissociation leads to 100% HNC + HCN). To estimate the rate constant values for these new reactions, density functional theory calculations were employed with the M06-2X functional and the aug-cc-pVTZ (hereafter AVTZ) basis set using the Gaussian suite program.[58] To build the network, we first considered the reactions of HNCHCN with hydrogen atoms which are the most abundant reactive species in Titan's atmosphere. HNCHCN is predicted to react slowly with H as there is a notable barrier at the M06-2X/AVTZ level in good agreement with previous calculations.[54] We then consider the reactions of HNCHCN with N($^2$D) and protonation by H$_3^+$, HCNH$^+$ and C$_2$H$_5^+$. These reactions, which are assumed to be barrierless, are losses of HNCHCN involving similar fluxes to photodissociation. The major results of the photochemical model in terms of the simulated abundances as a function of altitude are shown in Figure 4.

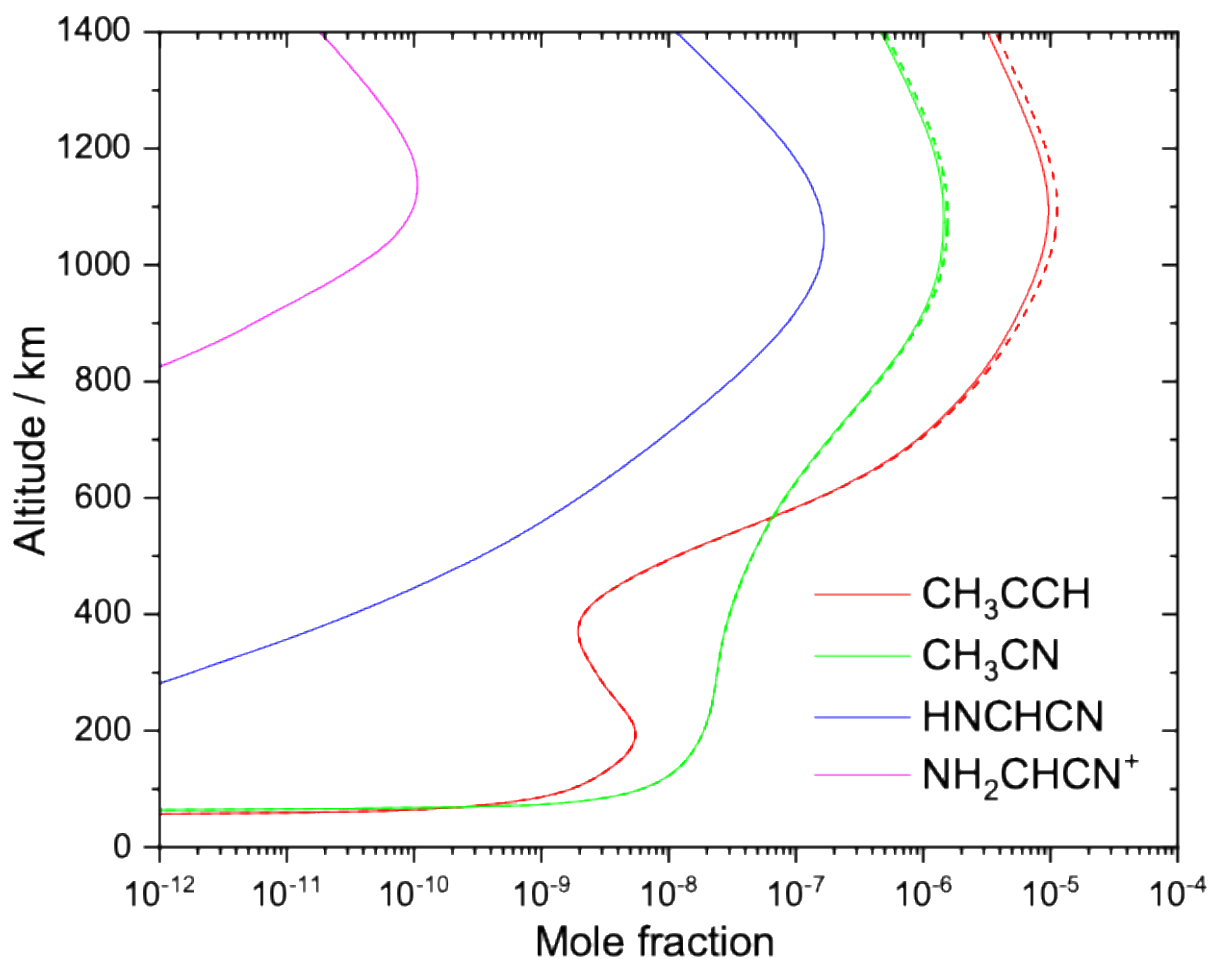

**Figure 4** Mole fraction profiles of CH₃CCH (red lines), CH₃CN (green lines), HNCHCN (blue line) and NH₂CHCN⁺ (purple line). Solid lines represent the simulations including reactions (4) and (5), dashed lines represent the simulations in the absence of these two reactions.

The introduction of the reaction N($^2$D) + CH$_3$CCH only leads to very minor changes in the model with a maximum decrease in simulated CH$_3$CCH abundances of around 15% around 1100 km. This arises because the flux of this reaction is low compared to the fluxes of the main reactions consuming N($^2$D) (N($^2$D) + C$_2$H$_2$ and C$_2$H$_4$ reactions) and CH$_3$CCH (photodissociation). Furthermore, this reaction does not lead to new products; species which are more efficiently produced by other reactions.

The situation is quite different for the N($^2$D) + CH$_3$CN reaction, despite a maximum decrease in modelled CH$_3$CN abundances of only 7% around 1100 km. Firstly, because the flux of this reaction is significant compared to the photodissociation flux of CH$_3$CN. This is not due to a larger reaction rate, as the rate constants are similar to those of the N($^2$D) + CH$_3$CCH reaction, but because the absorption of CH$_3$CN is shifted deeper into the UV range compared to that of CH$_3$CCH, which induces a much more effective shielding of CH$_3$CN by

$C_2H_2$ and $C_2H_4$ than for $CH_3CCH$. Secondly, this reaction produces a new species, HNCHCN. A related species, $NH_2CHCN^+$, is also formed due the proton transfer reactions of HNCHCN with $HCNH^+$ and $C_2H_5^+$ in Titan's ionosphere. HNCHCN, as a closed shell species is relatively unreactive, particularly with atomic hydrogen. Furthermore, it does not photodissociate very easily. This allows it to reach significant abundances, with a maximum of around 1e-7 relative to $N_2$ at around 1,000 km. Considering the uncertainties on the estimated loss processes for HNCHCN, we estimate at most an order of magnitude uncertainty on its simulated abundance. This abundance may eventually allow its detection through its rotational transitions. Indeed, HNCHCN has a strong dipole moment and has been detected using microwave spectroscopy towards Sagittarius B2(N). [59] HNCHCN is potentially important for the chemistry of Titan's atmosphere because it is formally the dimer of HCN. Its formation may be directly linked to the potential polymerization of HCN, which is suspected of playing a role in the formation of aerosols in Titan's atmosphere.[60] Furthermore, it has been suggested that HNCHCN can also be produced by the hydrogenation of $C_2N_2$.[61] This is also important information for the photochemistry of Titan's atmosphere, where $C_2N_2$ has been definitively detected.[62]

Given the potential importance HNCHCN to Titan's atmosphere, it is also interesting to quantify the possible contribution of both HNCHCN and $NH_2CHCN^+$ to m/z 54 (neutral spectrum) and 55 (ion spectrum) respectively detected by the Cassini INMS instrument.[63] According to these simulations, HNCHCN represents approximately 29 % of the total for neutral species contributing to mass 54 at its peak altitude (1045 km) with the other contributing species $CH_3CH_2CCH$ (61%), $CH_2CHCHCH_2$ (3%) and $C_2H_4CN$ (7%). According to the photochemical model results, for m/z 55, $NH_2CHCN^+$ accounts for only 6% of the total cations at its peak (1143 km) with the rest due to the presence of abundant $C_4H_7^+$ ions at this altitude.

## 6 Conclusions

This paper describes an experimental kinetic study of the $N(^2D) + CH_3CN$ and $N(^2D) + CH_3CCH$ reactions at room temperature and below. Considering the potential importance of both of these processes in the formation of nitrogen bearing organic molecules in planetary atmospheres, the new rate constants for these reactions were included in a photochemical model of Titan's atmosphere to test their effects on atmospheric species. Experimentally, a Laval nozzle or supersonic flow reactor was employed, allowing kinetic measurements to be performed over the 50-296 K range (75-296 K for the $N(^2D) + CH_3CN$ reaction). On the production side, $N(^2D)$ atoms were generated chemically rather than photolytically during the present work, via the $C(^3P) + NO \rightarrow N(^2D) + CO$ reaction. Here, the $C(^3P)$ atoms required for

N($^2$D) formation were produced by the 266 nm pulsed laser photolysis of CBr$_4$. On the detection side, N($^2$D) atoms were followed by pulsed laser induced fluorescence at 116.7 nm; a wavelength generated by the frequency doubling of a dye laser followed by the frequency tripling of the frequency doubled output in a rare gas cell. Both reactions are characterized by large rate constants displaying little or no temperature dependence, which contrasts with the estimates of these quantities in previous photochemical models where much smaller rate constants with large positive temperature dependences were used. On the photochemical modelling side, the rate constants and product channels of these reactions were updated according to the results of the present and recent experimental and theoretical studies, alongside the most important loss processes for the products themselves. Although, the inclusion of the N($^2$D) + CH$_3$CCH reaction has only a very minor influence on the overall chemistry, the N($^2$D) + CH$_3$CN reaction is predicted to generate HNCHCN as the major reaction product in the upper atmosphere of Titan. As HNCHCN is mostly unreactive due to its closed shell configuration and is also only photolyzed at short UV wavelengths, it reaches relatively large fractional abundances that may even be large enough to allow for its detection.

**Data availability**

Data for this article, including pseudo-first-order rate coefficients as a function of coreagent concentration are available at Zenodo at https://doi.org/10.5281/zenodo.17412644. Data supporting the photochemical modeling studies in this article have been included as part of the Supplementary Information.

**Acknowledgements**

K. M. H. acknowledges support from the French program ''Physique et Chimie du Milieu Interstellaire'' (PCMI) of the CNRS/INSU with the INC/INP co-funded by the CEA and CNES as well as funding from the ''Program National de Planétologie'' (PNP) of the CNRS/INSU.

**Electronic supplementary information file for "A Gas-Phase Kinetic Study of the N($^2$D) + CH$_3$CCH and N($^2$D) + CH$_3$CN Reactions"**

**Table S1 New/updated reactions added to the reaction network**

| Reaction | α | β | γ | F$_0$ | g | Comments and references |
|---|---|---|---|---|---|---|
| N($^2$D) + CH$_3$CCH → CH$_2$NH + C$_2$H | 5.9e-11 | 0.05 | 0 | 1.2 | 0 | Product channels taken from Mancini et al.[52] with the rate constant parameters derived in the present work |
| → CH$_3$ + HCCN | 4.0e-11 | 0.05 | 0 | 1.2 | 0 | |
| → C$_2$H$_3$CN + H | 2.3e-11 | 0.05 | 0 | 1.2 | 0 | |
| → c-CH$_2$CNCH+H | 5.8e-12 | 0.05 | 0 | 1.2 | 0 | |
| N($^2$D) + CH$_3$CN → HNCHCN + H | 1.2e-10 | 0 | 0 | 0.23 | 0 | Product channels taken from Mancini et al.[53] with the rate constant parameters derived in the present work |

Reaction rate are expressed as $k = \alpha \times (T/300)^\beta \times \exp(-\gamma/T)$ cm$^3$ s$^{-1}$
Uncertainties are expressed as $F(T) = F_0 \times \exp(g \times |1/T - 1/300|)$

**Photodissociation cross sections**

Photodissociation cross sections of HNCHCN were calculated using vertical excitation energies, transition dipole moments and oscillator strengths calculated at the EOM-CCSD/AVTZ level using Gaussian16.[57] The branching ratio of the HNCHCN photodissociation is taken to be equal to 100% towards HNC + HCN between 100 and 300 nm.

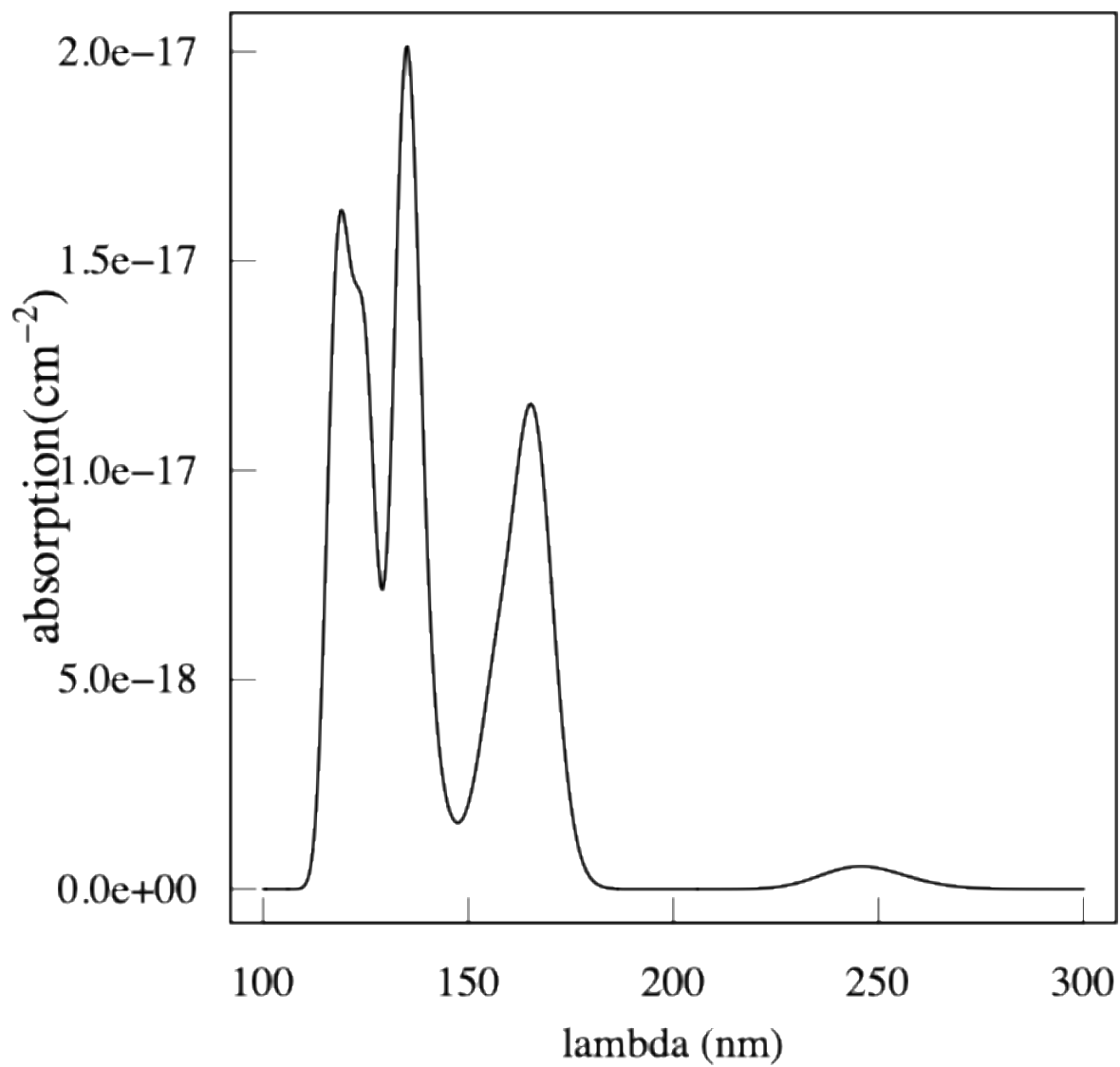

**Figure S1** Photodissociation cross sections of HNCHCN calculated at the EOM-CCSD/AVTZ level.